\documentclass[aps,prr,floatfix,preprintnumbers,altaffilletter,superscriptaddress,reprint,longbibliography,twocolumn]{revtex4-1}
\usepackage{graphicx,url,amssymb,amsmath,rotating,color,units,wasysym,epsfig,multirow,epstopdf, stackengine}
\usepackage{amsmath}
\usepackage{cancel}
\usepackage{physics}
\usepackage{appendix}
\usepackage{amssymb}
\usepackage[colorlinks,urlcolor=blue,citecolor=blue,linkcolor=blue]{hyperref}
\usepackage{soul,xcolor}
\usepackage[printonlyused,nolist]{acronym}
\usepackage{comment}
\usepackage{bm}
\usepackage[normalem]{ulem}

\newcommand{\pistroke}{
  \text{\protect\ooalign{\hidewidth\raisebox{-0.2ex}{--}\hidewidth\cr$\pi$\cr}}
}

\newcommand{\Lstroke}{%
  \text{\protect\ooalign{\hidewidth\raisebox{0.2ex}{--}\hidewidth\cr$\mathcal{L}$\cr}}%
}

\newcommand{\Mstroke}{%
  \text{\protect\ooalign{\hidewidth\raisebox{0.2ex}{--}\hidewidth\cr$M$\cr}}%
}

\newcommand{\Bstroke}{%
  \text{\protect\ooalign{\hidewidth\raisebox{0.5ex}{--}\hidewidth\cr$\mathcal{B}$\cr}}%
}

\newcommand{\chieff}{\chi_\textrm{eff}}
\renewcommand{\L}{\mathcal{L}}
\newcommand{\CIT}{\affiliation{Department of Physics, California Institute of Technology, Pasadena, California 91125, USA}}
\newcommand{\CITLab}{\affiliation{LIGO Laboratory, California Institute of Technology, Pasadena, California 91125, USA}}

\begin{document}

\title{Model exploration in gravitational-wave astronomy with the \\ maximum population likelihood}

\author{Ethan Payne}
\email{epayne@caltech.edu}
\CIT
\CITLab
\affiliation{School of Physics and Astronomy, Monash University, VIC 3800, Australia}
\affiliation{OzGrav: The ARC Centre of Excellence for Gravitational-Wave Discovery, Clayton, VIC 3800, Australia}

\author{Eric Thrane}
\email{eric.thrane@monash.edu}
\affiliation{School of Physics and Astronomy, Monash University, VIC 3800, Australia}
\affiliation{OzGrav: The ARC Centre of Excellence for Gravitational-Wave Discovery, Clayton, VIC 3800, Australia}

\begin{abstract}
Hierarchical Bayesian inference is an essential tool for studying the population properties of compact binaries with gravitational waves.
The basic premise is to infer the unknown prior distribution of binary black hole and/or neutron star parameters such component masses, spin vectors, and redshift.
These distributions shed light on the fate of massive stars, how and where binaries are assembled, and the evolution of the Universe over cosmic time. 
Hierarchical analyses model the binary black hole population using a prior distribution conditioned on hyper-parameters, which are inferred from the data.
However, a misspecified model can lead to faulty astrophysical inferences.
In this paper we answer the question: given some data, which prior distribution––from the set of all possible prior distributions––produces the largest possible population likelihood?
This distribution (which is not a true prior) is $\pistroke$ (pronounced ``pi stroke''), and the associated \textit{maximum population likelihood} is $\Lstroke$ (pronounced ``L stroke'').
The structure of $\pistroke$ is a linear superposition of delta functions, a result which follows from Carath{\'e}odory's theorem.
We show how $\pistroke$ and $\Lstroke$ can be used for model exploration/criticism.
We apply this $\Lstroke$ formalism to study the population of binary black hole mergers observed in LIGO--Virgo--KAGRA’s third Gravitational-Wave Transient Catalog.
Based on our results, we discuss possible improvements for gravitational-wave population models.
\end{abstract}

\maketitle

\section{Motivation}
Bayesian inference has become a mainstay of modern scientific data analysis as a means of analysing signals in noisy observations.
This procedure determines the posterior distributions for parameters given one or more model. 
In order to study the \textit{population properties} of a set of uncertain observations, a hierarchical Bayesian framework can be employed.
The basic idea is to model the population using a conditional prior $\pi(\theta | \Lambda, M)$, which describes, for example, the distribution of black hole masses $\{m_1, m_2\} \in \theta$ given some hyper-parameters $\Lambda$, which determine the shape of the prior distribution. 
Here, $M$ denotes the choice of model. 
One then carries out Bayesian inference using a ``population likelihood''
\begin{align}\label{eq:population_likelihood}
    {\cal L}(d | \Lambda, M) = \prod_i^N \frac{1}{\xi(\Lambda)}\int \dd\theta_i \, 
    {\cal L}(d_i|\theta_i) \pi(\theta_i | \Lambda, M) ,
\end{align}
where ${\cal L}(d_i | \theta_i)$ is the likelihood for data associated with event $i$ given parameters $\theta_i$, and $\xi(\Lambda)$ is the detected fraction for a choice of hyper-parameters.
Meanwhile, $N$ is the total number of observations.
For an overview of hierarchical modeling in gravitational-wave astronomy including selection effects, see Refs.~\cite{intro,Vitale2022,Mandel2019}.

The LIGO-Virgo-KAGRA (LVK) Collaboration's third gravitational-wave transient catalog (GWTC-3)~\cite{gwtc-3} contains the cumulative set of observations of $N=69$ confident binary black-hole mergers~\footnote{We adopt the threshold utilized in \citep{gwtc-3_pop} of a false-alarm-rate $< 1\,\textrm{yr}^{-1}$.} detected by the LVK~\citep{LIGO, Virgo, KAGRA}.
Additional detection candidates have been put forward by independent groups~\cite{Olsen_2022,Nitz_2021,Zackay_2021,Venumadhav_2020,Zackay_2019}.
Hierarchical inference is employed to study the population properties these merging binary black holes; see, e.g., Refs.~\cite{gwtc-1_pop,gwtc-2_pop,gwtc-3_pop,Roulet_2021, Farr_2017, mass, spin, Callister_2021, Fishbach_2022, Biscoveanu_2021, Biscoveanu_2022, Vitale_2017, Stevenson_2017, Miller2020, bbm, 2018_maya, Edelman_2022, Edelman2022_spline, golomb2022_spline}.
These analyses have revealed a number of exciting results, such as the surprising excess rate of mergers with a primary black hole mass of $\sim 35\,M_\odot$~\citep{gwtc-2_pop}, and the evolution of the binary merger rate with redshift~\citep{gwtc-3_pop}, to name just two.

However, Bayesian inference has its limitations.
One can use Eq.~\eqref{eq:population_likelihood} in order to infer the distribution of binary black hole parameters---\textit{given some model}; and one can compare the marginal likelihoods of two models to see which one better describes the data.
However, Bayesian inference does not tell us if any of the models we are using are suitable descriptions of the data. 
While all models for the distribution of binary black hole parameters are likely to be imperfect, some may be adequate for describing our current dataset \footnote{Here, we paraphrase the aphorism attributed to statistician, George Box: ``all models are wrong, but some are useful.''}.
When a model fails to capture some salient feature of the data, it is said to be ``misspecified'' \citep{wmf,Gelman}.
Some effort has been made to assess the suitability of gravitational-wave models, both qualitatively and quantitatively; see, e.g., \cite{gwtc-2_pop,gwtc-3_pop,wmf,Essick:2021:outliers}.
However, the idea of ``model criticism''---testing the suitability of Bayesian models---is still being developed within the context of gravitational-wave astronomy and beyond.

Hierarchical Bayesian inference studies often depend upon parametric models.
Modelers design parameterizations in order to capture the key features of the astrophysical distributions.
However, one must still worry about ``unknown unknowns''---features which do not occur to the modeler to add.
For example, recent studies~\cite{gwtc-2_pop,gwtc-3_pop, Callister2022, Tong2022} find a sub-population of binary black holes merge with spin vectors that are misaligned with respect to the orbital angular momentum axis. However, the degree to which the spins are misaligned might be model dependent. 
In Refs.~\cite{gwtc-2_pop,gwtc-3_pop, Callister2022}, the inferred minimum spin tilt is confidently $\gtrsim 90^\circ$. In contrast, Refs.~\cite{Roulet_2021,bbm,Tong2022} argue this signature could be due to a lack of flexibility in LVK models to account for a sub-population of black holes with negligible spin magnitude, finding support for misalignment at smaller minimum tilt angles. The inferred population distribution of spin misalignment has important consequences for understanding the formation channels of binary black-hole channels. 
This debate highlights how astrophysical inferences can be affected by model design. 

In order to help alleviate some of the issues arising from model misspecification in Bayesian inference, we present a framework for assessing the suitability of a model.
This framework is built around the concept of the \textit{maximum population likelihood} $\Lstroke$ (pronounced ``L stroke'')---the largest possible value of ${\cal L}(d|\Lambda)$ in Eq.~\eqref{eq:population_likelihood}, maximized over all possible choices of population model $\pi(\theta|\Lambda)$ \textit{independent of the choice of parameterization}.
The ``prior'' distribution, which yields this maximum is $\pistroke(\theta)$ (pronounced ``pi stroke'').
It is not a true prior because it is determined by the data.
The theory behind the maximization of population likelihoods has been studied previously in optimization and statistics literature~\citep{kiefer1956, simar1976,Laird1978,Bohning1982,Lindsay1983,Jiang2009}. 
This work is underpinned by Carath{\'e}odory's theorem \cite{Caratheodory} and the mathematics of convex hulls~\citep{Lindsay1983}.
However, its application to observational science has been somewhat limited as far as we can tell.

The $\Lstroke$ framework is useful for several reasons.
First, the numerical value of $\Lstroke$ is an upper bound on the population likelihood. 
We can compare the maximum likelihood for a specific model
\begin{equation}
\mathcal{L}_\textrm{max}(M) = \max_{\Lambda \sim p(\Lambda | d)}{\cal L}(d | \Lambda, M)
\end{equation}
to  $\Lstroke$.
Often in Bayesian model selection, the Bayesian evidence values (${\cal Z}_i$) of two hypotheses can be used to determine the extent to which one model is preferred over the other. 
A typical threshold chosen to rule out one model in favor of another is that $ \ln({\cal Z}_1/{\cal Z}_2) > 8$~\citep{jefferys1961}. 
In a similar vein, if $\ln(\Lstroke/\mathcal{L}_\textrm{max}(M)) \lesssim 8$, we can be sure the model $M$ is not badly misspecified since there is no second model $M'$ that can be written down with that will yield a statistically significant improvement.
We emphasize that a model which does not satisfy this condition is not necessarily misspecified.

Second, the $\Lstroke$ framework can be used to quantitatively assess if a model $M$ is misspecified.
By generating synthetic data from $M$, one can generate the expected distribution of $(\Lstroke, \mathcal{L}_\textrm{max}(M))$.
In this paper, we show how one can compare the observed values of $(\Lstroke, \mathcal{L}_\textrm{max}(M))$ to the expected distribution in order to determine the extent to which $M$ is misspecified---and the \textit{way} in which it is misspecified.

Third, the $\Lstroke$ framework can be used for ``model exploration''---providing clues of \textit{where} in parameter space unmodeled features might be lurking.
By comparing $\pistroke(\theta)$ with the prior from our phenomenological model $\pi(\theta|M)$, one can see if the phenomenological model is capturing key structure present in $\pistroke$ and use the comparison to design new models to test on forthcoming datasets.

The remainder of this paper is organized as follows.
In Sec.~\ref{sec:mpl}, we introduce the $\Lstroke$ formalism, illustrating key features with a simple toy model.
In Sec.~\ref{sec:crit}, we show how the formalism can be used for model criticism.
In Sec.~\ref{sec:applications}, we apply the formalism to study the population properties of merging binary black holes observed by the LVK. 
Our concluding remarks are presented in Sec.~\ref{sec:conc}.

\section{The maximum population likelihood \Lstroke}\label{sec:mpl}
\subsection{Preliminaries}
We begin with a brief review of Bayesian hierarchical inference with a parametric model.
Our starting point is the population likelihood (copied here from Eq.~\eqref{eq:population_likelihood}):
\begin{equation}\label{eq:margL}
        {\cal L}(d | \Lambda, M) = \prod_i^N \frac{1}{\xi(\Lambda)}\int \dd\theta_i \, 
    {\cal L}(d_i|\theta_i) \pi(\theta_i | \Lambda, M) .
\end{equation}
Here, $\L(d_i|\theta_i)$ is the likelihood of event-$i$ data $d_i$ given parameters $\theta_i$.
The quantity $\pi(\theta_i|\Lambda, M)$ is a conditional prior for $\theta_i$ given hyper-parameters for some population model $M$, which describes the shape of the prior distribution.
The term $\xi(\Lambda)$ accounts for selection effects; for example, high-mass systems are typically easier to detect than low-mass systems.
It is the detectable fraction of the population given the model given hyper-parameters $\Lambda$
\begin{equation}\label{eq:pdet}
    \xi(\Lambda) = \int \dd\theta \, p_\textrm{det}(\theta)\pi(\theta|\Lambda, M) .
\end{equation}
Here, $p_\textrm{det}(\theta)$ is the detection probability of an observation with parameters $\theta$. 

\subsection{The maximum population likelihood $\Lstroke$}~\label{subs:mpl}
The maximum population likelihood $\Lstroke$ is obtained by taking Eq.~\eqref{eq:margL} and maximizing over all possible prior distributions $\pi(\theta)$.
Thus, $\Lstroke$ is an upper bound (or supremum) on the set of likelihoods from all possible choices of models for $\pi(\theta)$ such that 
\begin{equation}\label{eq:mpldef1}
    \Lstroke \equiv \L({d}|\Mstroke) \geq \L({d} | \Lambda, M) ,
\end{equation}
for all models $M$.
The ``prior'' distribution that yields $\Lstroke$ is denoted
\begin{align}\label{eq:pistroke}
    \pistroke(\theta)
\end{align}
(pronounced ``pi stroke'').
It is not a true prior because the distribution which maximizes the population likelihood in Eq.~\eqref{eq:margL} depends on the data.
One should therefore refer to \pistroke as a pseudo-prior.
The associated model is denoted $\Mstroke$ (pronounced ``M stroke'').
Combining this notation into a single equation, we have
\begin{equation}\label{eq:mpldef3}
    \Lstroke \equiv \prod_{i=1}^N \frac{1}{\xi(\Mstroke)}\int \dd \theta_i\, \L(d_i|\theta_i)\pistroke(\theta_i) .
\end{equation}

\subsection{Calculating $\pistroke$: special cases}~\label{subs:struct}
Having introduced the concept of $\Lstroke$ and $\pistroke$, the natural next question is: given data $d$, how does one calculate these quantities?
Before answering this question, we study three special cases where we can work out $\pistroke$ from intuition.
This discussion will help sharpen our instincts for the more general solution that follows.
Readers looking to skip to the punchline may wish to skip this subsection.

\subsubsection{A single measurement}
For the first case, we consider a single measurement ($N=1$) with a unimodal likelihood function ${\cal L}(d |\theta)$, which is maximal when the parameter $\theta$ is equal to the maximum likelihood value $\widehat\theta$.
For the sake of simplicity, we ignore selection effects so that $\xi(\Mstroke)=1$.
In this case, $\Lstroke$ in Eq.~\eqref{eq:mpldef3} is clearly maximized if the prior support is entirely concentrated at $\widehat\theta$.
Thus, $\pistroke$ is a delta function
\begin{align}
    \pistroke(\theta) = \delta(\theta - \widehat\theta) ,
\end{align}
which yields
\begin{align}
    \Lstroke = & \int \dd\theta \, {\cal L}(d|\theta) \, 
    \delta(\theta - \widehat\theta) \nonumber\\
    = & {\cal L}(d|\widehat\theta) .
\end{align}
This result is intuitive: the prior that maximizes the population likelihood is the one that concentrates all its support at the maximum-likelihood value of $\theta$.

\subsubsection{$N$ signals in the high-SNR Limit}

For the second case, we consider a scenario in which the data consists of $N$ observations carried out in the high-SNR limit.
In this limit, the likelihood of the data for each measurement $d_i$ given some parameter $\theta$ approaches a delta function
\begin{align}
    {\cal L}(d_i | \theta_i) = \delta(\theta_i - \widehat\theta_i) ,
\end{align}
located at the maximum-likelihood value $\widehat\theta_i$.
We assume that each measurement is distinct so that no two maximum-likelihood values $\widehat\theta_i$ are exactly the same.
Again, for the sake of simplicity, we ignore selection effects so that $\xi(\Mstroke)=1$, though, the argument here holds even if we relax this assumption.
Equation~\eqref{eq:mpldef3} becomes
\begin{align}\label{eq:Lstroke_case2}
    \Lstroke = \prod_{i=1}^N \int \dd\theta_i \, 
    \delta(\theta_i - \widehat\theta_i) 
    \pistroke(\theta_i) .
\end{align}
The population likelihood is maximized when $\pistroke$ is a sum of delta functions peaking at the set of $\{\widehat\theta_i\}$:
\begin{align}\label{eq:pistroke_case2}
    \pistroke(\theta) = &
    \sum_{k=1}^N w_k \,
    \delta(\theta - \widehat\theta_k) \\
    w_k = & 1/N. \label{eq:weights} 
\end{align}
This solution for $\pistroke$ ensures that there is maximal prior support at every likelihood peak.
Obviously, the population likelihood is not maximized if any prior probability density is wasted to values of $\theta$ where all the likelihood functions are zero.
Choosing an equal weight for each delta function $w_i=1/N$ produces the largest possible population likelihood~\footnote{This is a well-known result known as the empirical distribution function~\citep{Laird1978}.}.

\begin{figure*}
    \centering
    \includegraphics[width=\linewidth]{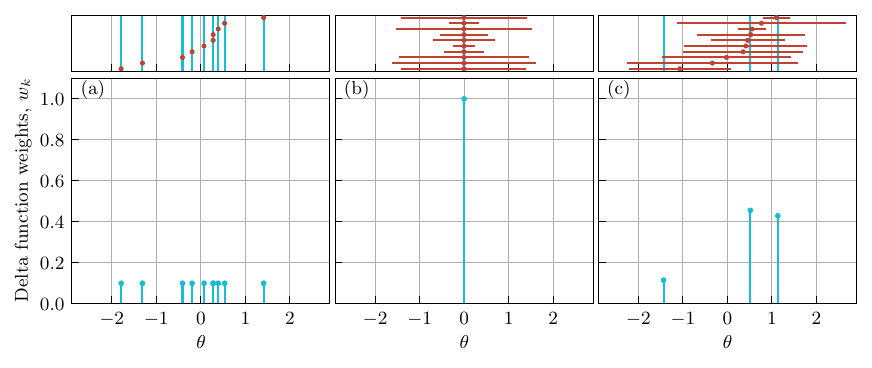}
    \caption{
    Examples of the distribution $\pistroke(\theta)$ described in Subsections~\ref{subs:mpl}-\ref{subs:general}.
    Each column represents a different dataset.
    The top-panel dots show the set of $N=10$ maximum-likelihood estimates $\{\widehat\theta_i\}$.
    The top-panel horizontal lines represent error bars; (in the first column they are too small to see), and the vertical lines (blue) indicate the inferred delta function locations.
    The bottom panels show the distribution of $\pistroke(\theta)$ associated with each data set.
    The left-hand column (a) represents data in the high-SNR limit so that the likelihood functions for each measurement approach delta functions (this is why the error bars are not visible).
    In this case, $\pistroke(\theta)$ consists of $N$ delta functions, each associated with one of the maximum likelihood points $\widehat\theta_i$.
    In the middle column (b), we are no longer in the high-SNR limit, but the maximum likelihood points are all assumed to be identical with $\widehat\theta_i = 0$.
    In this case, $\pistroke(\theta)$ consists of one delta function peaking at $\theta=0$.
    In the right-hand column (c), the data are not in the high-SNR limit, and each $\widehat\theta_i$ is random.
    In this case, $\pistroke(\theta)$ consists of $n=3$ delta functions, each with different heights.
    }
    \label{fig:1demo}
\end{figure*}

We illustrate this case in Fig.~\hyperref[fig:1demo]{1(a)} using high-SNR, toy-model data drawn from a mean-zero, unit-variance Gaussian distribution.
In the top-panel, we plot the set of $N=10$ maximum likelihood points $\{\widehat\theta_i\}$ and the position of the delta functions (blue).
In the lower panel, we ``plot'' the $\pistroke(\theta)$ for these ten data points.
We put the word ``plot'' in quotation marks because, technically, we are not plotting $\pistroke(\theta)$, which goes to infinity, but rather we are plotting the weights $w_k$ (Eq.~\eqref{eq:weights}), which allows us to see the relative weight given to each delta function---something that will prove useful below.
Throughout the paper, when we refer to plots of $\pistroke(\theta)$, it should be understood that we are actually plotting \textit{representations} of $\pistroke(\theta)$ using the weights $w_k$.
Finally, note that each peak in the distribution of $\pistroke(\theta)$ matches up with one of the maximum likelihood points in the upper panel.

\subsubsection{$N$ identical measurements}
For the third case, we consider a set of $N$ observations.
This time, we do not assume the high-SNR limit, but we assume that every measurement has the same maximum-likelihood value of $\widehat\theta$.
This case is highly contrived---one does not typically work with multiple identical measurements---but the example is nonetheless helpful for illustrative purposes.
In this case, the integral in Eq.~\eqref{eq:mpldef3} is maximized when the prior support is entirely concentrated at $\widehat\theta$ (where all of the likelihood functions peak), so that $\pistroke$ is a single delta function:
\begin{equation}
    \pistroke(\theta) = \delta(\theta - \widehat\theta) ,
\end{equation}
while
\begin{equation}
    \Lstroke = \prod_{i=1}^N\L(d_i|\widehat\theta).
\end{equation}
This scenario is demonstrated in Fig.~\hyperref[fig:1demo]{1(b)}.
The top panel shows the set of $N=10$ maximum-likelihood points $\{\widehat\theta_i\}$, all with the same value.
The horizontal lines represent the error bars for each measurement, which we draw from a uniform distribution on the interval $(0.01,1)$.
In the lower panel, we plot $\pistroke(\theta)$ for these ten data points.
This time, since every measurement is identical, $\pistroke(\theta)$ is a single delta function peaking at $\theta=0$.

From these three examples, we observe a pattern: in each case, $\pistroke(\theta)$ can be written as a weighted sum of delta functions.
Indeed, it has been proven that this is in fact the case~\citep{kiefer1956, simar1976,Laird1978,Bohning1982,Lindsay1983,Jiang2009}.
We refer readers interested in an explanation of the delta function structure of $\pistroke$ to Appendix~\ref{app:proof}, where we summarize the key concepts surrounding the proof outlined in Ref.~\citep{Lindsay1983} using the mathematics of convex hulls.
We do not reproduce the proof in its entirety, but rather we use visualisations to explain how it works with $N=2$ observations, before providing a qualitative explanation for how it generalizes to arbitrary values of $N$.
We explore this general structure and the consequences thereof in the next subsection.

\subsection{The general form of $\pistroke$}\label{subs:general}
We proceed with the knowledge that Eq.~\eqref{eq:mpldef3} is true in general, regardless of the form of the likelihood ${\cal L}(d | \theta)$ and the selection effect term $p_\text{det}(\theta)$.
\textit{For any set of observations, $\pistroke(\theta)$ is always of the form,}
\begin{equation}\label{eq:delta}
    \pistroke(\theta) = \sum_{k=1}^n w_k\,\delta(\theta - \theta_k),
\end{equation}
\textit{where $w_k$ are weights which sum to unity}
\begin{equation}\label{eq:constr}
    \sum_{k=1}^n w_k = 1 .
\end{equation}
The number of delta function is always less than or equal to the number of measurements and the solution is unique in all but the most pathological of cases (e.g., multimodal distributions with regions of equivalent maximum likelihoods) so that
\begin{align}
    n \leq N .
\end{align}
The ratio
\begin{align}\label{eq:informativeness}
    \mathcal{I} \equiv n/N ,
\end{align}
is a measure of the ``informativeness'' of the data.
It compares the typical likelihood width to the scatter in the astrophysical distribution.
In the high-SNR limit, $\mathcal{I}=1$, since a delta function is required for every data point (see Fig.~\hyperref[fig:1demo]{1(a)}).
The other limiting case is, $\mathcal{I}=1/N$, which happens when the likelihood for each measurement completely overlaps (see Fig.~\hyperref[fig:1demo]{1(b)}).

Using this insight into the structure of $\pistroke(\theta)$, we now consider a variation on the toy-model problems discussed in the earlier subsections.
In particular, we consider finite-SNR data drawn from our Gaussian, toy-model distribution.
Using Eqs.~(\ref{eq:delta}-\ref{eq:constr}) as an ansatz, we calculate $\pistroke(\theta)$ for $N=10$ random data points.
The maximum likelihood values $\widehat\theta_i$ are drawn from a mean-zero, unit-variance Gaussian and the error bars are drawn from a uniform distribution on the interval $(0.01, 1)$.
The results of this calculation are shown in Fig.~\hyperref[fig:1demo]{1(c)}.
The top panel shows the data, represented by the maximum-likelihood values $\{\widehat\theta_i\}$, which are arranged from bottom to top in increasing order.
The horizontal lines show the uncertainty for each measurement and the vertical blue lines indicate the positions of the delta functions.
In the bottom panel, we show $\pistroke(\theta)$ for this dataset.
It consists of just $n=3$ delta functions of varying heights ($\mathcal{I}=0.3$).
The exact weights, locations, and number of delta functions are not obvious; we obtain them numerically by maximising Eq.~\eqref{eq:delta} subject to Eq.~\eqref{eq:constr} using the ``combined'' method described below in Subsection~\ref{subs:comp}.
Comparing the red data points with error bars to the turquoise representation of $\pistroke(\theta)$, one can see that every data point can be plausibly associated with at least one of the delta functions.

Given the form of $\pistroke(\theta)$ described by Eq.~\eqref{eq:delta}, we can write down a general expression for $\Lstroke$:
\begin{equation}\label{eq:delta_mpl}
    \Lstroke = \prod_{i=1}^N 
    \frac{1}{\xi(\Mstroke)} \sum_{k=1}^n w_k \,
    \L(d_i|\theta_k) ,
\end{equation}
where 
\begin{equation}\label{eq:delta_xi}
    \xi(\Mstroke) = \sum_{k=1}^n w_k \, 
    p_\textrm{det}(\theta_k) .
\end{equation}
Given Eqs.~\eqref{eq:delta_mpl} and~\eqref{eq:delta_xi}, the problem of calculating $\Lstroke, \pistroke$ reduces to the problem of simply finding the locations and weights of $n$ delta functions.
In Section~\ref{subs:comp}, we explore three different approaches to this problem.

\subsection{Computing \pistroke}\label{subs:comp}
In this subsection, we consider three techniques that can be applied to compute $\Lstroke, \pistroke$: optimization, iterative grid, and stochastic methods.
We show that a combined approach, which uses a grid-based approach to guess a solution, which is subsequently refined through optimization performs the best out of the algorithms we tried.
Meanwhile, the stochastic approach allows us to illustrate the existence of the delta function structure proven in Ref.~\cite{Lindsay1983}, but with minimal assumptions.

\begin{figure}
    \centering
    \includegraphics[width=\linewidth]{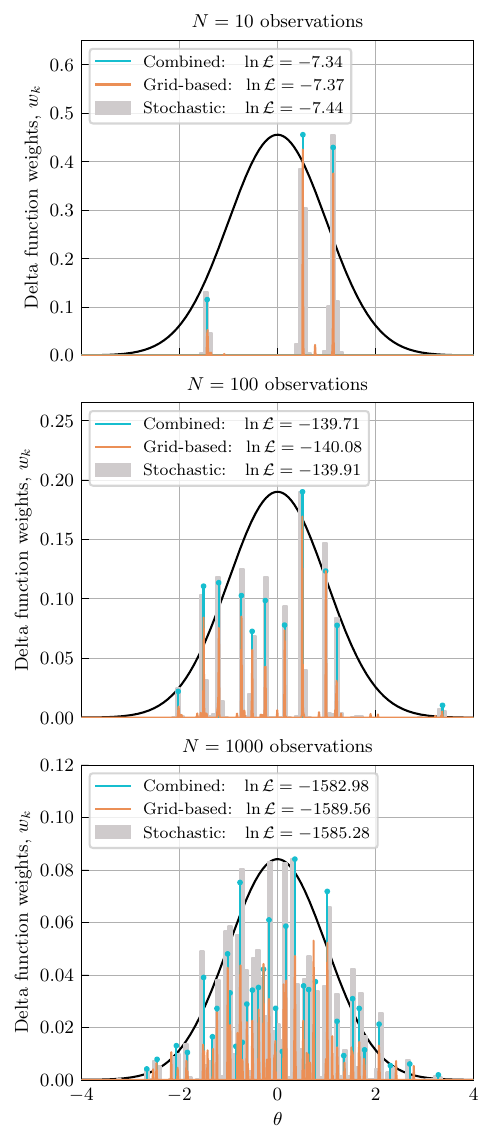}
    \caption{
    Demonstration of different methods for calculating $\pistroke, \Lstroke$.
    Each panel shows the results for a different number of measurements with $N=10$ on top, $N=100$ in the middle, and $N=1000$ on the bottom.
    The black distribution is the true distribution $\pi(\theta)$ used to generate the data.
    The colored spikes show the reconstructed distribution $\pistroke(\theta)$ as determined by different methods.
    Cyan is for the ``combined'' technique, which uses the iterative grid to obtain a first guess that is refined with the optimization method.
    Meanwhile, orange is for the grid-based technique by itself and gray is for the stochastic method.
    }
    \label{fig:2demo}
\end{figure}

\subsubsection{Optimization}~\label{subsubs:opt}
The first approach we consider is to use an optimization algorithm subject to the constraint in Eq.~\eqref{eq:constr}~\footnote{In theory, the constraint condition does not need to be enforced during the analysis. 
The normalization appears in the selection function term and \pistroke. 
However, since any multiple of the weights (without normalization) would produce an identical likelihood, many numerical optimization methods can falter at these likelihood ``plateaus''. 
Therefore, we enforce the constraint to ensure a more robust analysis.}.
We use {\tt \sc Scipy}'s \texttt{trust-constr} optimization implementation~\citep{2020SciPy, Conn2000}. 
We find this approach fails to find the correct global maximum of Eq.~\eqref{eq:delta_mpl} once the number of peaks $n$ becomes large. 
However, this issue can be resolved if a sufficiently close guess to the true shape of $\pistroke(\theta)$ can be made.
Fortunately, the iterative-grid approach can be used to supply this initial guess.

\subsubsection{Iterative grid}~\label{subsubs:grid}

The second approach we consider is to iteratively place delta functions on a fixed grid.
There are two steps: the greedy addition of many delta functions, and the removal of no-longer-useful delta functions. 
In the first step, we first attempt to place a delta function with a fixed height at each grid point and evaluate Eq.~\eqref{eq:delta_mpl} (with appropriate normalization of the distribution). 
We determine which of all possible delta function additions produces the highest population likelihood. We then vary the height of this delta function between zero and twice the initial height in order to obtain an updated guess for $\pistroke(\theta)$. 
The addition of delta functions is repeated, reducing the initial height by a factor at each iteration. 
After many iterations, we then attempt to remove no-longer-useful delta functions to further increase the population likelihood. 
We repeat this procedure five times, iteratively adding 30 delta functions with varying heights at each iteration. 
After these iterations, $\Lstroke$ is usually well-converged for the problems we are studying. 
In some iterations, this procedure adds support to preexisting delta functions.
This is how the approach ``corrects'' under-supported delta functions. 

This method has a significant advantage over generic constrained optimization techniques as the procedure does not require the optimization of individual parameters governing the delta functions through the $\{\theta_k,w_k\}$ space.
However, we find that this method is improved by pairing it with optimization. 
The most accurate optimization of the maximum population likelihood and structure of the distribution occurs when we utilize grid-based approximation to inform the starting location and weights for the constrained optimization. 
This allows for the grid-based approximation to find the region of parameter space where $\Lstroke$ is nearly maximal. 
The constrained optimization then purifies the delta function structure and slightly increases the maximum population likelihood. 
The \textit{combined} method is used for all the maximum population likelihood computations in Sec.~\ref{sec:applications}. 

\subsubsection{Stochastic construction}\label{subsubs:stoch}
Our final approach is to stochastically generate samples for $\pistroke(\theta)$, which are accepted/rejected depending on whether the new samples increases the population likelihood.
This is a form of importance sampling in which an arbitrary ``proposal distribution'' is used to generate proposal samples.
When a proposal sample is generated, we add it to a list of previously accepted points and evaluate $\Lstroke$ as a Monte Carlo integral,
\begin{equation}
    \Lstroke = \prod_{i=1}^N \frac{1}{\xi(\Mstroke)}\Big\langle \L(d_i|\theta_i)\Big\rangle_{\theta_i \sim \pistroke(\theta_i)},
\end{equation}
where 
\begin{equation}
    \xi(\Mstroke) = \Big\langle p_\textrm{det}(\theta)\Big\rangle_{\theta \sim \pistroke(\theta)} .
\end{equation}
Here, the angled brackets indicate averaging over the samples. 
If the addition of the new sample increases $\Lstroke$, we retain the sample in the list of samples from $\pistroke$. As the process is repeated, the set of samples produces an ever-improving representation of $\pistroke$. 

This method can be extended to employ an additional burn-in phase and/or a thinning phase to ensure more rapid convergence by removing unfavorable samples that sometimes get accepted early on before the distribution is well-converged. 
While this approach converges more slowly than the other two methods, \textit{it does not employ any assumptions about the structure of the distribution}. 
Thus, this method can be used to validate the structure put forward in Eqs.~(\ref{eq:delta_mpl}-\ref{eq:delta_xi}), that $\pistroke(\theta)$ is a sum of delta functions.

\subsubsection{Numerical study}
We demonstrate each method using our Gaussian, toy-model distribution described in the last subsection: true maximum likelihood values $\widehat\theta_i$ drawn from zero-mean, unit-variance Gaussian with error bars drawn from a uniform distribution on the interval (0.01, 1).
The observed maximum likelihood values are then shifted from the true value by an offset generated from each individual observation's uncertainty.
The results of this demonstration are compiled in Fig.~\ref{fig:2demo}.
The three panels of Fig.~\ref{fig:2demo} represent tests performed with $N=10$, $100$, and $1000$ observations.
In each panel, the black curve represents the true distribution $\pi(\theta)$.
The colored spikes illustrate different numerical solutions for $\pistroke(\theta)$: cyan is the ``combined'' approach, which uses the iterative grid to obtain an initial guess that is subsequently refined using the optimization method.
Meanwhile, orange represents the iterative grid approach by itself.
For the grid-based approach we run 30 iterations of adding peaks with variable but decreasing weights, before repeating this process an additional ten times. Finally, gray represents the stochastic approach.
For the stochastic method, we generate 3000 samples with 1000 samples for burn-in. 

\begin{figure}
    \centering
    \includegraphics[width=\linewidth]{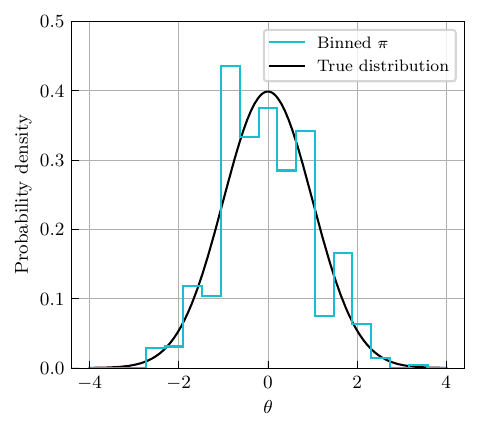}
    \caption{
    Comparison between a binned representation of $\pistroke$ as computed for the toy model data set with $N=1000$ observations and the true underlying population distribution. 
    This representation more clearly shows that $\pistroke$ is approaching the true distribution in the limit of many observations. 
    }
    \label{fig:binned_comp}
\end{figure}

We see that the combined approach better estimates $\Lstroke$ relative to the other techniques considered \footnote{A method is ``better'' if it yields a larger value of $\Lstroke$ than another approach.}.
We observe that, as $N$ increases, $\pistroke(\theta)$ increasingly resembles the true Gaussian distribution $\pi(\theta)$ (shown in Fig.~\ref{fig:2demo} as a black curve).
To illustrate this more clearly, we take the inferred delta function locations from the $N=1000$ ``combined'' result in Fig.~\ref{fig:2demo} and compute the weighted histogram. This result is directly compared to the true distribution in Fig.~\ref{fig:binned_comp}, from which we see that indeed the inferred distribution is (albeit slowly) approaching the true distribution. 
\textit{We conjecture that, in general, $\pistroke(\theta)$ approaches the true distribution in the infinite-data limit:}
\begin{align}
    \lim_{N \rightarrow \infty} \pistroke(\theta) \rightarrow \pi_\textrm{true}(\theta) .
\end{align}

\subsubsection{Computational challenges}
Before continuing, we discuss two computational challenges.
First, we note that the examples illustrative above are all one-dimensional.
The discussion above generalizes to $\geq 2$ dimensions; $\pistroke(\theta)$ is still a sum of delta functions in $\geq 2$ dimensions.
However, it becomes increasingly challenging to determine the location and height of these peaks in higher dimensions.
Furthermore, by increasing the dimensionality of the problem, constructing continuous representations of the individual-event likelihoods and the detection probability, $p_\textrm{det}(\theta)$, becomes increasing difficult. 
Recent developments in using Gaussian mixture models to produce continuous representations of these distributions might alleviate these concerns~\cite{Talbot:2020oeu, Golomb:2021tll}.
Second, even if we stay in one dimension, the computational cost of calculating $\pistroke, \Lstroke$ grows with $N$~\footnote{For the results in Fig.~\ref{fig:2demo}, the computation time of the ``combined'' approach was the following: $10$ observations required only $5.3$ seconds, $100$ observations required $65$ seconds and $1000$ observations required $2780$ seconds.
Generally,  more data tends to require more delta functions (each with a location and a height), meaning the computational difficulty grows with $N$.}.

\section{Model criticism with $\Lstroke$}~\label{sec:crit}
In this section, we show how the $\Lstroke$ formalism can be used to determine if a model $M$ is an adequate description of data.
The first step is to generate synthetic datasets based on the posterior distribution for the model hyper-parameters $p(\Lambda | d)$.
For each data set, we calculate the maximum population likelihood $\Lstroke$ (Eq.~\eqref{eq:mpldef3}) as well as the maximum likelihood for $M$, which we denote
\begin{align}
    \mathcal{L}_\textrm{max}(M) = \max_{\Lambda \sim p(\Lambda | d)}{\cal L}(d | \Lambda, M)
\end{align}
where $\L(d|\Lambda, M)$ is the population likelihood defined in Eq.~\eqref{eq:margL}.
In this way we can estimate 
\begin{align}
    p(\Lstroke, \L_\textrm{max}(M)) ,
\end{align}
the joint distribution for $\Lstroke$ and $\L_\textrm{max}(M)$ given model $M$.
By comparing the \textit{measured} values of $(\Lstroke, \L_\textrm{max}(M))$, to this distribution of \textit{expected} values, one can see if the dataset is typical of what one would expect given $M$.
If the measured values of $(\Lstroke, \L_\textrm{max}(M))$ are atypical, one can conclude that $M$ is misspecified.
Moreover, one may determine the nature of the misspecification by noting the location of the observed value of $(\Lstroke, \L_\textrm{max}(M))$ relative to the typical values of $(\Lstroke, \L_\textrm{max}(M))$.
This is best illustrated with an example.

In our example, we imagine that an observer measures $N=100$ values of some parameter $\theta$.
Their model $M$ for the distribution of $\theta$ consists of a Gaussian distribution with mean $\mu=0$ and width $\sigma=1$:
\begin{align}\label{eq:M}
    \pi(\theta|M) \sim {\cal N}(\mu=0, \sigma=1) .
\end{align}
However, their model may be misspecified so that $\theta$ is not really distributed according to $M$.
We consider five ``possible worlds'' \footnote{We borrow the language of ``possible worlds'' from the philosopher, David Lewis, who invokes them in his account of counterfactuals and necessity~\cite{Lewis}.}, one in which the observer's model is correctly specified and four in which it is not.
Each world is assigned a color: 
\begin{itemize}
    \item Black: model is correctly specified $(\mu=0,\sigma=1)$.
    \item Purple: model is too wide because the true distribution is $(\mu=0,\sigma=0.6)$.
    \item Blue: model is too narrow because the true distribution is $(\mu=0,\sigma=1.4)$.
    \item Salmon: model is shifted to one side because the true distribution is $(\mu=1,\sigma=1)$.
    \item Yellow: model is too wide \textit{and} shifted to one side because the true distribution is $(\mu=0.8, \sigma=0.6)$.
\end{itemize}
We create ten mock datasets for each of the five possible worlds (black, purple, blue, salmon, and yellow) and 5000 mock datasets from the model $M$ (grey contours).
For each dataset, we compute $(\Lstroke, \L_\textrm{max}(M))$---always using model $M$  (Eq.~\ref{eq:M}) even if the data are generated according to, say, the blue-world distribution.
This is because we are studying the case where our observer might apply a misspecified model.

\begin{figure*}
    \centering
    \includegraphics[width=\linewidth]{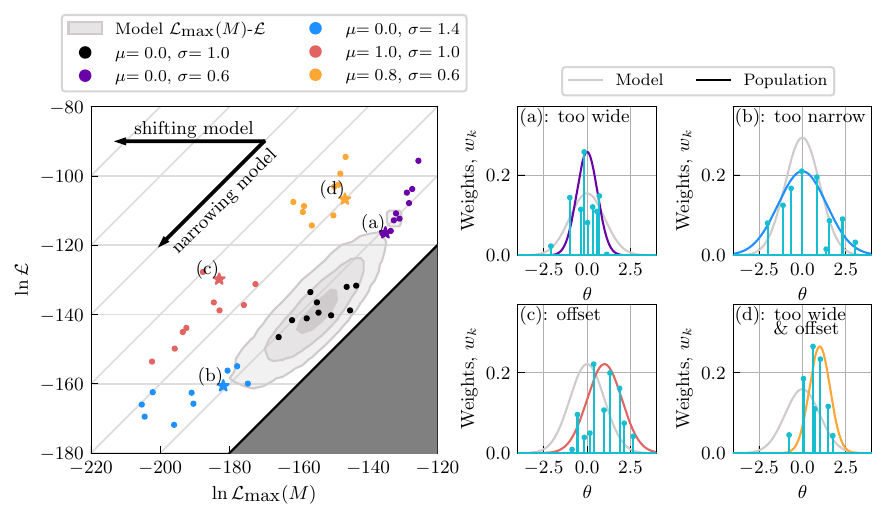}
    \caption{
    An illustration of model criticism with the $\Lstroke$ formalism.
    In the left-hand panel, we plot $(\Lstroke, \L_\textrm{max}(M))$ for five different underlying populations (each with ten different realizations), analyzed a toy-model with a mean of $\mu=0$ and standard deviation $\sigma=1$.
    Each population is represented by a different color.
    The gray contours show the 1, 2, and 3-sigma credible intervals for the expected distribution of $p(\Lstroke, \L_\textrm{max}(M))$ from the toy-model.
    By comparing the measured values of $(\Lstroke, \L_\textrm{max}(M))$ from an observed population to the expected distribution from our choice of model, one may determine if the dataset is typical of what one would expect given the model.
    If the measured values of $(\Lstroke, \L_\textrm{max}(M))$ fall outside these intervals, one may conclude that the toy-model is misspecified (does not accurately model the data). 
    Moreover, the location of a point on this plot relative to the expected distribution, conveys information about the way in which a model is misspecified.
    The right-hand panel shows the toy-model (grey), the true population distribution for the starred and labeled datapoint (a-d), and the respective $\pistroke$ for the observed data (turquoise). 
    This demonstrates that shifts away from the expected distribution (left-hand panel; grey) in $(\Lstroke, \L_\textrm{max}(M))$ can be visually identifiable to the reconstruction of $\pistroke$. }
    \label{fig:dist}
\end{figure*}

The results are shown in Fig.~\ref{fig:dist}.
The vertical axis is $\ln\Lstroke$ while the horizontal axis is $\ln \L_\textrm{max}(M)$.
The dark-grey region in the bottom-right corner is forbidden since $\Lstroke \geq \L_\textrm{max}(M)$ by construction.
The grey contours show the one, two, and three-sigma contours for the expected distribution from the model.
Only the black world datasets are consistent with the expected distribution, as the model is correctly specified in the black world. 
The colored dots, meanwhile, show ten random realizations of ($\ln\Lstroke, \ln \L_\textrm{max}(M)$) in colored worlds where the model is misspecified in various ways.
This is fundamentally different from a typical Bayesian inference plot where the data are fixed and the model is varied.
Here, the model is fixed to $M$ (Eq.~\ref{eq:M}), and we consider different datasets, which may or may not be misspecified depending on the world of our observer.

When the model $M$ is sufficiently misspecified with respect to the true distribution, it becomes unlikely for our observer to obtain values of $(\Lstroke, \L_\textrm{max}(M))$ that reside within the expected three-sigma interval---a sign of misspecification.
Interestingly, the different colored dots cluster in different regions.
For example, in the world where the model $M$ is too broad (purple), the dots cluster above-right of the gray contours.
In the world where the model $M$ is shifted away from the true peak (salmon), the dots cluster to the left of the gray contours.
By studying \textit{where} one's observed values of $(\Lstroke, \L_\textrm{max}(M))$ fall on this diagram, one can gain some insight into the way in which one's model is misspecified.
This example focuses on relatively simple forms of misspecification involving the mean and variance.
Other forms of misspecification (e.g., involving skewness and kurtosis) are, of course possible as well.
Given all the ways that a model can be misspecified, the ``shifting model'' / ``narrowing model'' arrows on Fig.~\ref{fig:dist} should be taken as rule-of-thumb signposts.

In practice, it is computationally challenging to create plots like Fig.~\ref{fig:dist} for population studies in gravitational-wave astronomy.
While it is easy to create mock datasets, it is time-consuming to calculate individual-event likelihoods for one dataset, let alone thousands.
There may be workarounds.
We discuss this possibility in greater detail below.

\section{Application to gravitational-wave astronomy}\label{sec:applications}
In this section, we apply the $\Lstroke$ formalism to results from gravitational-wave astronomy to stress-test models for the population of merging binary black holes.
We analyze data from the second gravitational-wave transient catalog (GWTC-3) \cite{gwtc-3, gwtc3_release}, which includes 69 confidently detected binary black hole mergers with false alarm rates $<\unit[1]{yr^{-1}}$.
To ensure similarity to the GWTC-3 LVK population analysis~\cite{gwtc-3_pop, gwtc3_pop_release}, we utilize the same individual-event posterior samples --- constructed from equally weighted samples generated from effective-one-body ({\tt \sc SEOBNRv3}~\citep{Pan_2013, Taracchini_2013}, {\tt \sc SEOBNRv4PHM}~\citep{Bohe_2016gbl, Ossokine_2020kjp}) and phenomenological ({\tt \sc IMRPhenomPv2}~\citep{Hannam_2013}, {\tt \sc IMRPhenomXPHM}~\citep{Pratten_2020ceb}) waveform results (see~\cite{gwtc-3_pop} for more details).
To construct the lower-dimensional individual-event likelihoods, we utilize the same samples while marginalizing over all other ``\textit{nuisance}'' parameters. For these ``nuisance'' parameters, we chose the distributions associated with the \textit{maximum a posteriori} hyper-parameters from the LVK's GWTC-3 population analysis with the {\tt \sc Power Law+Peak}-{\tt \sc Default}-{\tt \sc Power Law} model~\citep{gwtc-3_pop}.

We divide out the sampling prior to convert the one-dimensional posterior to a likelihood.
The likelihood normalization is computed using the Bayesian evidence of each event. 
The normalization is not important for the calculation of $\pistroke$, but it affects the misspecification tests demonstrated in Sec.~\ref{sec:gwtc3crit}.
We calculate the hyperparmeter distributions and $\L_\textrm{max}(M)$ using \textsc{GWPopulation} \cite{gw_population}, which employs \textsc{Bilby} \cite{bilby,bilby_gwtc1} and {\tt \sc Dynesty}~\citep{speagle2020dynesty}. 
We utilize the combined injection set from Ref.~\cite{gwtc3_sens_release} to compute the estimated detectable fraction of binary black-hole mergers over the first three observing runs. 

\subsection{Model inspiration through visual inspection}~\label{subs:inspire}
One straightforward application of the $\Lstroke$ formalism is to visually compare the reconstructed population distribution (obtained using a phenomenological model) with $\pistroke(\theta)$. 
By comparing these two distributions, it is possible to see which features in the phenomenological model reconstruction are due to prior assumptions, which features are due to real trends in the data, and which features might be missing from the phenomenological model.
Formally, we compare $\pistroke(\theta)$ to the population predictive distribution (PPD)
\begin{align}
    \text{PPD}(\theta | d, M) =
    \int \dd\Lambda \, 
    p(\Lambda | d) \pi(\theta | \Lambda, M) ,
\end{align}
which describes the astrophysical distribution of $\theta$ given a phenomenological model $M$ with hyper-parameters $\Lambda$.

\begin{figure}
    \centering
    \includegraphics[width=\linewidth]{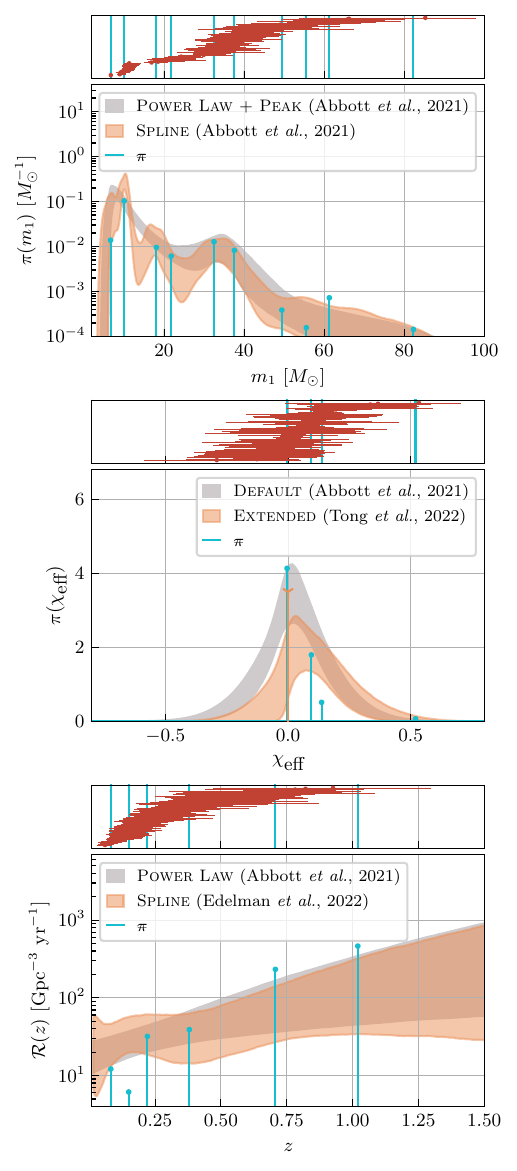}
    \caption{
    Population predictive distributions ($90\%$ credibility) and $\pistroke$ for primary the black-hole mass ($m_1$; \textit{top}), effective inspiral parameter ($\chieff$; \textit{middle}), and redshift ($z$; \textit{bottom}) distributions.
    For the redshift, we divide by the evolution of the comoving volume and time delay as a function of redshift to plot the merger rate, $\mathcal{R}(z)$. 
    Comparison of the different models with $\pistroke$ highlights which features are present in the data and which are due to assumptions in the model.
    } 
    \label{fig:gwmpl}
\end{figure}

In Fig.~\ref{fig:gwmpl}, we present $\pistroke(\theta)$ with the PPDs from the LVK analysis of GWTC-3 \cite{gwtc-3_pop, gwtc3_pop_release} for source-frame primary mass $m_1$ (top), the effective inspiral spin parameter $\chieff$ (middle), and redshift $z$ (bottom).
Each row contains two sub-panels; the small upper panel shows the maximum-likelihood estimate for each gravitational-wave event and the $90\%$ confidence interval while the larger lower panel compares $\pistroke$ with the PPD.
The PPD is plotted as a thick band to show the 90\% credibility region at each value of $\theta$.

We first turn our attention to the primary mass distribution in the top row.
There are $M=10$ delta function peaks, implying an informativeness of $\mathcal{I}=0.15$ (see Eq.~\eqref{eq:informativeness}).
This result is computed in 169.3 seconds.
The gray band is the \textsc{Power Law + Peak} model from \cite{mass} while the orange band is a (more flexible) semi-parametric power-law-spline model denoted \textsc{Spline} from \cite{Edelman_2022}.
The agreement between $\pistroke$ and the two PPDs is striking, with cyan spikes closely matching several of the features in both models including the turn-over at low masses near $\approx 12 M_\odot$ and the bump at $30 M_\odot$.
Furthermore, we see that $\pistroke$ also recovers some of the finer detail features found only by the \textsc{Spline} model.
In particular, the shift in the low-mass peak and the dips in posterior support at $\sim16\,M_\odot$ and $\sim25\,M_\odot$ are present in the structure of $\pistroke$.
Based on our visual inspection, it appears that current models are capturing much if not all of the structure present in $\pistroke$.

Turning our attention to the middle row, we study the distribution of effective inspiral spin parameter \cite{Damour2001},
\begin{align}\label{eq:chieff}
    \chieff \equiv \frac{\chi_1 \cos\theta_1 + q \chi_2 \cos\theta_2}{1+q} ,
\end{align}
which measures the mass-weighted black hole spin projected along the orbital angular momentum~\footnote{In Eq.~\eqref{eq:chieff}, $q\equiv m_2/m_1$ is mass ratio, $\chi_{1,2}$ are the dimensionless black hole spins, and $\theta_{1,2}$ are the spin vector tilt angles relative to the orbital angular momentum.}.
This time, only $n=4$ delta function spikes are required to fit the data ($\mathcal{I} = 0.06$), showing how much harder it is to measure $\chieff$ than $m_1$.
Computing $\pistroke(\chieff)$ requires 71.3 seconds. The quicker computation time is likely a result of the lower number of delta functions required.
In gray, we plot the PPD for the \textsc{Default} model from Refs.~\cite{gwtc-2_pop,gwtc-3_pop}, which draws on work from Refs.~\cite{spin,Wysocki2019}.
In orange we plot the PPD for the \textsc{Extended} model from Refs.~\cite{bbm,Tong2022}, which only analyse 68 binary black-hole events in the population due to data quality concerns regarding one event~\citep{Payne:2022spz}. 
To plot the {\tt \sc Extended model} results, which incorporates a delta function at $\chieff=0$, we plot the $90\%$ interval for the delta function height, $\delta$, multiplied by the same scale factor as $\pistroke$. 
The continuous contribution to the {\tt \sc Extended} model is then scaled by the ratio of the PPD evaluated at only the non-zero $\chieff$ $\pistroke$ delta functions to the previously computed scaling. 

The data-driven $\pistroke$ includes a delta function at $\chieff \approx 0$ and three smaller peaks in the $\chieff > 0$ region, but no peaks with $\chieff < 0$.
The lack of support for $\chieff<0$ is in contrast to Refs~\cite{gwtc-2_pop, gwtc-3_pop}, which find support for a sub-population of binary black holes with $\chieff < 0$.
The strong delta function at $\chieff=0$ lends support to the argument put forward in Refs.~\cite{Miller2020,Roulet_2021,bbm} that the data can be well-modeled with a sub-population of ``non-spinning'' $\chieff=0$ binaries, even if there is not strong statistical support for the existence of such a peak \cite{Callister2022,Mould2022,Tong2022}.
However, our visual comparison suggests that the \textsc{Extended} model may over-predict the abundance of binaries with $\chieff \approx 0.3$. 
Moreover, we note that the distribution of $\chieff=0$ appears to also be consistent with a smooth, one-sided distribution, maximal at $\chieff=0$, and slowly decaying at larger positive values of $\chieff=0$---that is, a single population.

Turning our attention to the bottom row of Fig.~\ref{fig:gwmpl}, we consider the case of redshift.
For this parameter, $n=6$ ($\mathcal{I} = 0.09$), and takes $116$ seconds to compute.
Here we plot the merger rate as a function of redshift, $\mathcal{R}(z)$ by dividing the posterior predictive distribution by the PPD by the evolution of the comoving volume and time delay with respect to redshift. 
The merger rate is more commonly utilized for interpreting the redshift evolution. 
The $\pistroke$ distribution fits a decrease in the merger rate at a redshift of $z\sim0.13$. 
While we caution that $\pistroke$ is purely data-informed, and such a feature might diminish with additional observations, the \textsc{Power Law} model utilized in Refs.~\cite{gwtc-2_pop,gwtc-3_pop} lacks the flexibility to resolve such a feature. 
Comparing our results to Ref.~\citep{Edelman2022_spline}, we observe that $\pistroke$ is qualitatively different from the ``non-parametric'' model~\footnote{Ref.~\citep{Edelman2022_spline}'s spline model is probably better described as ``ultra-parameterized.''} used in that paper. 
Our best guess is that the reconstruction from Ref.~\citep{Edelman2022_spline} is reasonable, and that the different features in $\pistroke$ are due to noise fluctuations, though, it is possible that the smooth spline structure imposed by the \citep{Edelman2022_spline} model is misspecified or that the prior on ``knot location'' is somehow subtly influencing the fit.
As more gravitational-wave observations are made, finer structure may emerge in the redshift evolution of the binary merger rate. 
These differences between the parametric reconstructions and $\pistroke$ might present the first hints of such structure.
We suggest that future redshift models include additional flexibility to study the possibility of a deficit of mergers in the nearby Universe.

By using the iterative ``grid-based'' method (without further constrained optimization), we also demonstrate the computation of a two-dimensional $\pistroke$ distribution.
In particular, we study the joint distribution of mass ratio $q$ and effective spin inspiral parameter $\chieff$. 
Recent studies have explored the possibility of astrophysical correlations between $q$ and $\chieff$~\citep{callisterQXF2021, adamcewicz2022unequal, gwtc-3_pop}, finding an anticorrelation, i.e. more unequal mass systems typically possess a effective spin inspiral parameter. 
The presence of an anticorrelation in the $q$-$\chieff$ distribution has implications for the formation environments of binary black holes. 
Ref.~\citep{mckernan2022}, for example, propose that such an anticorrelation could be due to assembly of binary black holes in active galactic nuclei.

\begin{figure*}
    \centering
    \includegraphics[width=\linewidth]{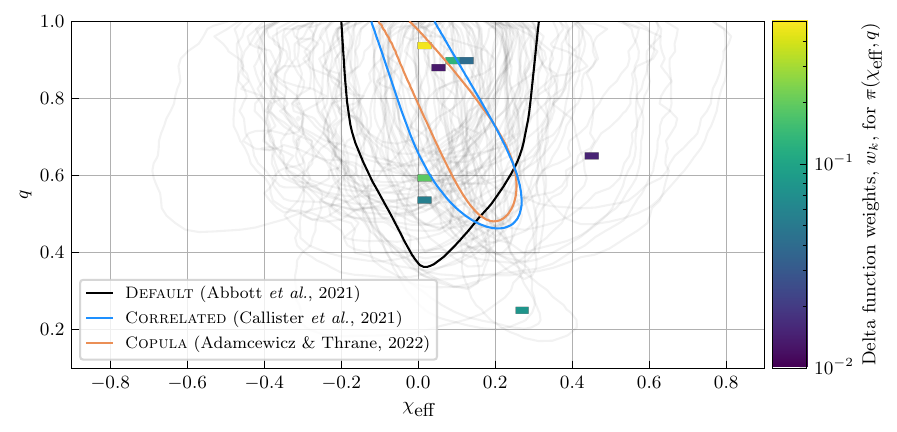}
    \caption{The joint distribution $\pistroke(q, \chieff)$ represented by eight colored pixels.
    The pixel color is related to the delta-function weight.
    The purely data-derived $\pistroke$ can be compared to the 90\% contours of \textit{maximum a posteriori} distribution estimates for three specific models.
    The black curve shows the reconstructed population given the \textsc{Default} model from Ref.~\cite{gwtc-3_pop} (which does not allow for correlation) while the blue and orange curves show the reconstructed population given by the \textsc{Correlated} model from Ref.~\cite{callisterQXF2021} and the {\tt \sc Copula} model from Ref.~\citep{adamcewicz2022unequal}, respectively.
    The grey contours correspond to the 90\% credible intervals of the 69 events in GWTC-3~\citep{gwtc-3, gwtc-3_pop}. 
}
    \label{fig:qchieff}
\end{figure*}

In Fig.~\ref{fig:qchieff} we plot $\pistroke(q, \chi_\text{eff})$ as eight colored pixels. 
It is easier to digest this \pistroke plot than the superposition of single-event, 90\% credible intervals for all 69 events (gray).
In order to compare \pistroke to recent models, we plot the 90\% contours of \textit{maximum a posteriori} distribution estimates for the {\tt \sc Default} model in Ref.~\citep{gwtc-3_pop} which assumes no correlation (black curve), the {\tt \sc Correlated} model from Ref.~\citep{callisterQXF2021} (blue curve) and the {\tt \sc Copula} model from Ref.~\citep{adamcewicz2022unequal}. 
From visual examination of \pistroke, it is clear that the anticorrelation identified in Ref.~\cite{callisterQXF2021} is based on actual features in the data: the pixels corresponding to the delta functions $\pistroke$ are consistent with anticorrelation between $(q, \chieff)$.
However, \pistroke is also consistent with they hypothesis that there are separate sub-populations located at different regions in the $q$-$\chieff$ space (an instance of Simpson's reversal \cite{Simpson}).

\subsection{Upper bounds on population model likelihoods}
In Table~\ref{tab:mpl_bounds} we report the difference in natural log likelihood comparing the various population models to the maximum population likelihood $\Lstroke$:
\begin{align}\label{eq:Bstroke}
    \ln\Bstroke \equiv \ln \Lstroke - \ln \L_\textrm{max}(M)  .
\end{align}
The $\ln\Bstroke$ values in Table~\ref{tab:mpl_bounds} measure the fit of population models relative to the best possible fit.
Motivated by the typical threshold for model selection in terms of Bayes factors~\citep{jefferys1961}, a value of $\ln\Bstroke\lesssim 8$ indicates that the population model is very close to the maximum population likelihood~\cite{intro}, which would imply that the fit cannot be dramatically improved.
A large value of $\ln\Bstroke$ by itself does not imply that a model is ``wrong'' or unsuitable to describe the data, but it does quantify the extent to which an alternative model can in-principle improve over the current offerings.

Returning to Table~\ref{tab:mpl_bounds}, the \textsc{Power Law + Peak} model for $m_1$ shows the most potential room for improvement.
This may be due to structure identified using the \textsc{Spline} model, which is missing from the less flexible \textsc{Power Law + Peak}.
However, the $m_1$ measurements are also the most informative in Table~\ref{tab:mpl_bounds} (with the largest value of $\mathcal{I}$).
With more information, it is probably easier to concoct an \textit{a posteriori} model with a large population likelihood that explains various features in the distribution of $m_1$ through over-fitting.
The \textsc{Default} and \textsc{Extended} spin models both exhibit $\ln\Bstroke<8$, which implies that neither model can be unequivocally ruled out, though, the \textsc{Extended} model provides a somewhat better fit with a natural log likelihood difference of $4.17$.
We also note that the $\chieff$ and $z$ observations are noticeably less informative, and simultaneously the associated values of $\L_\textrm{max}(M)$ are closer to $\Lstroke$. 
This might indicate that, while there are features present in $\pistroke$ that are present in the data, they are not statistically significant. 

\begin{table}
    \centering
    \begin{tabular}{c|c|c|c}
    \hline\hline
         Parameter & $\mathcal{I}$ & Model & $\ln\Bstroke$ \\\hline
         $m_1$ & 0.15 & \textsc{Power Law + Peak} & $14.89$ \\
         & & \textsc{Spline} & 6.66\\\hline
         $\chieff$ & 0.06 & \textsc{Default} & $7.70$ \\
         & & \textsc{Extended} & 3.53\\\hline
         $z$ & 0.09 & \textsc{Power Law} & $8.93$ \\
         & & \textsc{Spline} & $6.59$ \\\hline\hline
    \end{tabular}
    \caption{
    The performance of different population models relative to the $\Mstroke$.
    The quantity $\Bstroke$ (Eq.~\eqref{eq:Bstroke}) is a measure of the population likelihood of each model relative the maximum possible population likelihood $\Lstroke$.
    The ``informativeness'' $\mathcal{I}$ (Eq.~\eqref{eq:informativeness}) is a measure of the information available about the distribution of each parameter.
    }
    \label{tab:mpl_bounds}
\end{table}

\subsection{Model criticism in gravitational-wave astronomy}~\label{sec:gwtc3crit}

It would be interesting to make a version of the left-hand panel of Fig.~\ref{fig:dist} using the population models from gravitational-wave astronomy discussed in the previous subsection.
Unfortunately, this is quite computationally difficult.
First, we would need to run single-event parameter estimation of $N\approx 69$ events drawn from a random realization of the population fit to the observed gravitational-wave events. 
This needs to be repeated ${\cal O}(1000)$ times to produce the refined contours as those shown in the toy-model example (Fig.~\ref{fig:dist}). 
However, as an initial demonstration, we generate three simulated catalogs of 69 events using three draws from the {\tt \sc Power Law + Peak - Default - Power Law} hyperposterior informed by observations from GWTC-3~\citep{gwtc-3_pop}. 
These simulated observations were produced with injections of the \textsc{IMRPhenomXPHM}~\citep{Pratten_2020ceb} waveform into simulated Gaussian noise colored by the power spectral density from the first half of the third LVK observing run.

We then run Bayesian hierarchical inference to determine the posterior predictive distributions from the parameterized model. 
Using the posterior predictive distributions, following the calculation undertaken for the collection of real gravitational-wave observations, we produce the one-dimensional marginal likelihoods which are then used to compute $\Lstroke$ and ${\cal L}_\textrm{max}(M)$.
Unlike in Sec.~\ref{fig:dist}, where enough simulated catalogs are produced to construct an expected distribution in the $(\Lstroke, \L_\textrm{max}(M))$ plane, here we are required to model and fit the distribution. 
We employ Bayesian inference and a simple multivariate Gaussian distribution model to estimate the structure in the expected $(\Lstroke, \L_\textrm{max}(M))$ distribution. 
We use a Wishart prior on the covariance matrix~\citep{chung2015weakly}.
We use the posterior predictive distribution of fitted Gaussian distributions to estimate whether the models utilized in Ref.~\citep{gwtc-3_pop} are inadequate for the observations.

\begin{figure*}
    \centering
    \includegraphics[width=\linewidth]{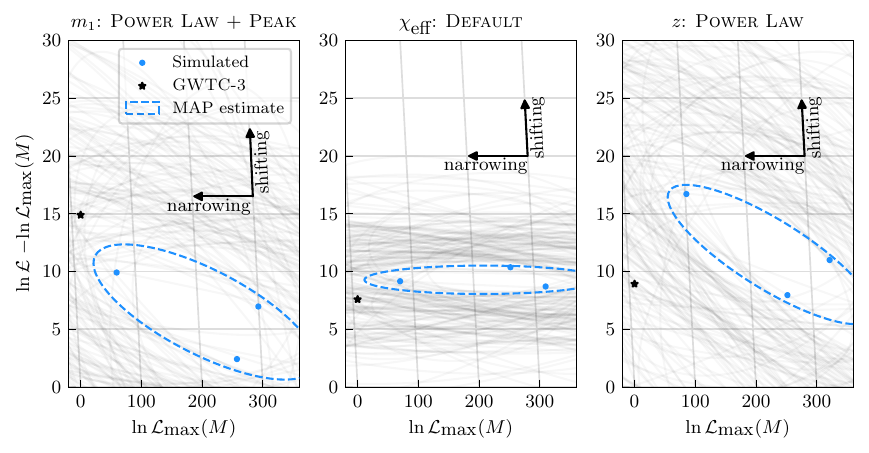}
    \caption{
    Demonstration of the $(\Lstroke, \L_\textrm{max}(M))$ model misspecification test for three parameterized models used in Ref.~\citep{gwtc-3_pop}. 
    Due to the limited number of simulated gravitational-wave catalogs, we model the expected distribution $p(\Lstroke, \L_\textrm{max}(M))$ as a multivariate Gaussian distribution and infer the possible mean and covariance matrix from the three simulated values (blue). 
    The grey ellipses correspond to the $3\sigma$ confidence intervals for $100$ different realizations of the possible distribution. 
    The dashed blue ellipses correspond to the \textit{maximum a posteriori} (MAP) predictive distributions. The inferred values of $(\Lstroke, \L_\textrm{max}(M))$ from the 69 events in GWTC-3 are shown by the black point.
    The likelihoods are normalized by the maximum likelihood inferred from the GWTC-3 model.
    From the inferred ellipses, we can conclude that there is a possibility that some or all models used are inadequate for the observations.
    Further studies with larger simulated catalogs are required to truly determine whether these models are misspecified.}
    \label{fig:gwtc3empdist}
\end{figure*}

The results are shown in Fig.~\ref{fig:gwtc3empdist} for the primary black-hole mass, effective inspiral parameter, and redshift. The blue dots correspond to the three simulated gravitational-wave catalogs, whereas the black star corresponds to the observed values from GWTC-3. 
The gray ellipses are $3\sigma$ intervals for $(\Lstroke, {\cal L}_\text{max}(M))$, each associated with a different realisation of our Gaussian fit.
(The large amount of scatter is due to the fact that we are attempting to fit a Gaussian to just three points.)
The dashed blue curve corresponds to the \textit{maximum a posteriori} (MAP) estimate. 
The value of $\L_\textrm{max}(M)$ has been normalized to the value found for GWTC-3. 
The inferred points in $(\Lstroke, \L_\textrm{max}(M))$ for GWTC-3 typically reside beyond the $3\sigma$ confidence interval, which we use as our criteria for misspecification.

We calculate a $p$-value for each panel, which quantifies the probability of observing the GWTC-3 values for $(\Lstroke, \L_\textrm{max}(M))$ given our fit; small $p$-values are indicative of misspecification.
For the \textsc{Power Law + Peak} primary black-hole mass model is misspecified we find $p=46\%$, for the \textsc{Default} $\chieff$ model we find $p=31\%$, and for the redshift \textsc{Power Law} model we find $p=8\%$. 
None of the models we consider are clearly ruled out as misspecified, as the sensitivity of this test is somewhat hamstrung by the small number of simulated catalogs.
It would not surprise us if a more aggressive follow-up study ${\cal O}(1000)$ simulations identified one or more models as more obviously misspecified.

One important caveat to these results is that the overall normalization of the likelihood depends on the computation of the individual observation Bayesian evidences. 
With stark differences between the analyses made in Refs.~\citep{gwtc-3, gwtc-3_pop}, it is difficult to accurately emulate the correct overall normalization of the likelihood. 
This globally impacts in the scale of ${\cal L}_\textrm{max}(M)$ for the simulated catalog -- potentially shifting the distributions closer or further from the inferred GWTC-3 result. 
In addition, the robustness of the evidences computed within Ref.~\citep{gwtc-3} are not guaranteed (see e.g. Ref.~\citep{Callister2022}).

There are a number of solutions to address the computational cost of this analysis.
While probably not realistic in the near future, it may be possible to represent the likelihood functions of simulated events using a Fisher matrix approximation, which would speed up the calculation significantly.
However, verifying that this approximation produces adequately estimates for $\Lstroke, \L_\textrm{max}(M)$ could remain a challenge.
Another possibility worthy of investigation is the idea that the distribution of $\Lstroke, \L_\textrm{max}(M)$ might have some quasi-universal properties.
If it can be shown that a large class of problems produce a similarly-shaped distribution of $\Lstroke, \L_\textrm{max}(M)$, perhaps a relatively small number of simulations can be used to work out the shape of $p(\Lstroke, \L_\textrm{max}(M))$.
We leave this for future work.
Perhaps most promising are efforts to speed up inference with various machine learning schemes; see, e.g., Ref.~\cite{Dax}.
As these tools become more reliable, it may become possible to estimate $(\Lstroke, \L_\textrm{max}(M))$ in a matter of seconds, which would in turn enable precision tests of misspecification.

\section{Conclusion}~\label{sec:conc}
The $\Lstroke$ formalism provides a useful lens through which to view population studies in gravitational-wave astronomy.
It provides an upper bound on the Bayesian evidence for population models, $\Lstroke$.
The associated pseudo-prior distribution $\pistroke$ is a sum of delta functions.
The $\pistroke$ distribution can be used to see which features in a reconstructed distribution are model-dependent, and which are genuinely present in the data.
The $\pistroke$ distribution can also draw attention to features in the data that are not fit by current models, providing a tool for the design of new models.
Finally, the $\Lstroke$ formalism can be used to determine if a model is misspecified, by comparing the values of $(\Lstroke, \L_\textrm{max}(M))$ to the expected distribution of these quantities given the model $M$.
This comparison can be made quantitatively with a $p$-value.
And, by comparing the measured values of $(\Lstroke, \L_\textrm{max}(M))$ to the distribution expected given the model, it is possible to see the way in which the model is misspecified.
Constructing a distribution of $\Lstroke, \L_\textrm{max}(M)$ may be computationally prohibitive in gravitational-wave astronomy, though, future work is required to investigate simplifying assumptions that might bring down the cost.

While we have introduced the $\Lstroke$ formalism within the context of gravitational-wave astronomy, the framework is general, and we expect it can be applied to a broad range of problems in astronomy and beyond where one seeks to infer the distribution of parameters $\theta$ with potentially unreliable hierarchical models.

\section*{Acknowledgements}
We thank Katerina Chatziioannou, Colm Talbot, Isaac Legred, Isobel Romero-Shaw, and Paul Lasky for insightful discussions about the $\Lstroke$ formalism. 
We thank Rory Smith for input on early discussions regarding using $\Lstroke$ for model mispecification tests. 
We are grateful to Jacob Golomb for discussions focused on computing a continuous representation of the detection probability for gravitational-wave astronomy.
We are indebted to Bernard Whiting for important discussions regarding the convex hull formulation of population distributions. 
We thank Tom Callister for comments on an early version of the manuscript. 

This material is based upon work supported by NSF's LIGO Laboratory which is a major facility fully funded by the National Science Foundation.
This research has made use of data, software and/or web tools obtained from the Gravitational Wave Open Science Center (https://www.gw-openscience.org), a service of LIGO Laboratory, the LIGO Scientific Collaboration and the Virgo Collaboration.
Virgo is funded by the French Centre National de Recherche Scientifique (CNRS), the Italian Istituto Nazionale della Fisica Nucleare (INFN) and the Dutch Nikhef, with contributions by Polish and Hungarian institutes.
The authors are grateful for computational resources provided by the LIGO Laboratory and supported by National Science Foundation Grants PHY-0757058 and PHY-0823459.
This paper carries LIGO Document Number \#P2200309.
E.T. is supported through Australian Research Council (ARC) Centre of Excellence CE170100004 and ARC DP230103088.

\appendix

\section{Outline of $\pistroke$ structure proof~\label{app:proof}}
\subsection{Overview}
\begin{figure*}
    \centering
    \includegraphics[width=0.86\linewidth]{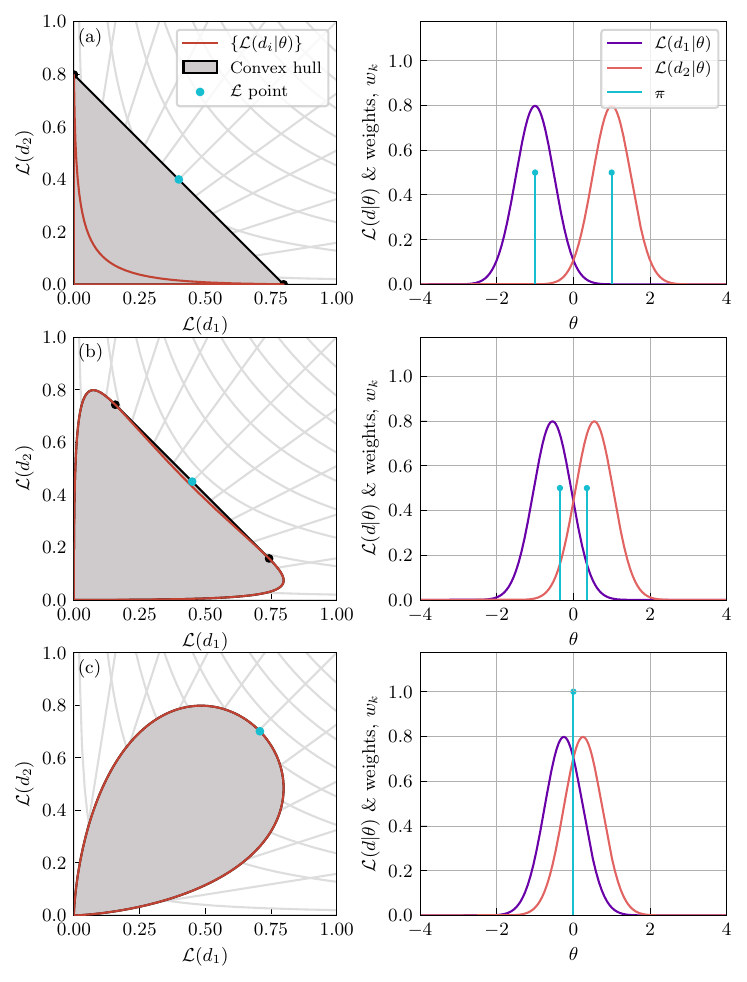}
    \caption{
    Visual illustrations of the proof in Ref.~\citep{Lindsay1983}. 
    The left-hand column panels show the atomic likelihood vectors (red), the convex hull produced from the red curve (grey with black outline), and the cyan point on the convex-hull boundary with the maximum population likelihood \Lstroke. 
    The black points correspond to the points from the set of atomic likelihood vectors which generate the maximum population likelihood. 
    The right-hand column panels show three examples of $N=2$ single-event likelihood functions (purple and red).
    The distribution of $\pistroke$ is indicated with one or more cyan spikes.
    These spikes correspond to the $\Lstroke$ solution (cyan dot) in the corresponding left-hand panel. 
    In (a), the two single-event likelihoods are mostly disjoint and so two delta functions are required to maximize the population likelihood (cf. Fig. 1 in Ref.~\citep{Lindsay1983}). 
    As the two single-event likelihoods begin to overlap further, these two delta functions move closer together as shown in (b). 
    Moving the single-event likelihoods closer still, the set of atomic likelihood vectors becomes the boundary of the convex hull, at which point only one delta function is required to maximize the likelihood as shown in (c).  
    }
    \label{fig:hull}
\end{figure*}

In this appendix we outline the basic ideas underpinning the proof from Ref.~\cite{Lindsay1983} by Lindsay that \pistroke consists of a sum of $\leq N$ delta functions:
\begin{align}
    \pistroke(\theta) = \sum_{k=1}^n w_k \, 
    \delta(\theta - \theta_k) .
\end{align}
Our aim is to provide readers with a qualitative understanding.
To this end, we consider a simple example of $N=2$ measurements, each characterized by a Gaussian likelihood functions.
Our example measurements are depicted in the right-hand column of Fig.~\ref{fig:hull}, which shows two single-event likelihoods (one in purple, the other in red), both conditioned on some parameter $\theta$.
In each row of Fig.~\ref{fig:hull}, we vary the separation of these two single-event likelihood functions relative to their width: far apart in the top row, becoming closer together in the two subsequent rows.
We show below how \pistroke consists of either one or two delta functions, depending on this relative separation and explain how this generalizes to $N>2$.

Lindsay's proof relies on the mathematics of \textit{convex hulls}, geometric shapes which can be defined in arbitrarily high dimensions.
If one draws a line between any two points on a convex hull, all the points on that line are also part of the hull.
(The gray shaded regions in the left-hand column of Fig.~\ref{fig:hull} are all examples of convex hulls.)
Convex hulls are often used in optimization problems with constraints where the optimal solution occurs on the boundary of the hull, which is determined by the constraints.
In Lindsay's proof, the relevant constraint equation is the unitarity of the $\pistroke(\theta)$:
\begin{align}
    \int d\theta \, \pistroke(\theta) = 1 .
\end{align}
The unitarity constraint means that the form of $\pistroke(\theta)$ that maximizes the population likelihood exists on the boundary of a complex hull.

\subsection{A geometric picture}
For the sake of simplicity, we ignore the impact of the selection function~\footnote{The selection function term, $p_\textrm{det}(\theta)$, can be absorbed into the prior to determine $\pistroke$ on the observed population before correcting the detection probability afterwards.}.
We represent the observations using what Lindsay refers to as an \textit{atomic likelihood vector},
\begin{equation}
    \bm{L}(\widehat\theta) \equiv \{\mathcal{L}(d_1|\widehat\theta), \mathcal{L}(d_2|\widehat\theta), ..., \mathcal{L}(d_N|\widehat\theta)\} .
\end{equation}
Each element of this vector is a single-event likelihood marginalised over a delta-function prior peaking at $\widehat\theta$:
\begin{align}
    {\cal L}(d_i | \widehat\theta) = & 
    \int d\theta_i \, 
    {\cal L}(d_i | \theta_i) \, 
    \delta(\theta - \widehat\theta) .
\end{align}
This allows us to represent the problem in an abstract $N$-dimensional likelihood space.
The left-hand column of Fig.~\ref{fig:hull} provides a visualization of such a two-dimensional atomic likelihood vector space.
Scanning over all possible values of $\widehat\theta$ traces out the red curve in the atomic likelihood vector space, which represents all possible values of the atomic likelihood vector $\bm{L}(\theta)$.
By varying $\widehat\theta$, we can make an individual element of the atomic likelihood vector large, but doing may make other elements of the vector small as we see in the top row with widely separated single-event likelihood functions.

The weighted sum of atomic likelihood vectors 
\begin{align}\label{eq:bmL}
    \bm{L}(\vec{w}) = \sum_k w_k \, \bm{L}(\widehat\theta_k)
\end{align}
yields a vector of likelihoods with elements
\begin{align}
    {\cal L}(d_i | \vec{w}) = 
    \sum_k w_k \, 
    {\cal L}(d_i | \widehat\theta_k) ,
\end{align}
corresponding to the marginal likelihood given a prior of delta functions
\begin{align}
    \pi(\theta) = \sum_k  w_k \,
    \delta(\theta - \widehat\theta_k) ,
\end{align}
where 
\begin{align}
    \sum_k w_k = 1 .
\end{align}
This means we can construct more general \textit{marginal likelihood vectors} with a linear combination of atomic vectors.
Furthermore, in the continuum limit, \textit{any} prior can be used to \textit{marginalize} over the atomic likelihood vectors. 
Elements of the marginal likelihood vector in the continuum limit take the form,
\begin{equation}
    {\cal L}(d_i | M) = \int d\widehat\theta_i \, {\cal L}(d_i | \widehat\theta_i) \, \pi(\widehat\theta_i|M).
\end{equation}

Let us consider again the $N=2$ example illustrated in Fig.~\ref{fig:hull}.
If we pick any two points on the red curve, each corresponding to some value of $\widehat\theta$, which we denote $A$ and $B$, we can define two basis vectors: $\hat{e}_A$ and $\hat{e}_B$.
The linear combinations of these two basis vectors forms a line connecting $A$ and $B$.
All of the points along this line represent likelihood vectors constructed from $N=2$ delta functions.
By connecting together every possible pair of points on the red atomic likelihood points, we map out the gray region---the convex hull.
Every possible marginal likelihood vector (for \textit{any} choice of prior) is part of the hull.
That is, the set of all possible summations is the convex hull and is a representation of all possible probability distributions in the likelihood space. 
This result is profound---our original problem is reduced from an infinite set of possible population distributions to a closed region in an $N$-dimensional likelihood space. 
The construction of the convex hull is unique \cite{Lindsay1983}, except in pathological cases further discussed in Sec.~\ref{sapp:pathological}. 

Now that we have studied the geometry of the atomic likelihood vector space, we ask the question: what point in our convex hull corresponds to the maximum population likelihood?
The population likelihood can be written as a product of the marginal likelihood vector elements:
\begin{align}
    {\cal L}_\text{pop}(\vec{d} |M) = \prod_{i=1}^N {\cal L}(d_i|M) .\label{eq:app_poplikelihood}
\end{align}
In $N=2$ dimensions, we can fix ${\cal L}_\text{pop}(\vec{d})$ and identify hyperbolic curves of the form
\begin{align}\label{eq:hyperbolic}
    {\cal L}(d_2) = {\cal L}(\vec{d}) / 
    {\cal L}(d_1) ,
\end{align}
represented in the left-hand column of Fig.~\ref{fig:hull} by gray curves.
All the points on one of these curves have the same population likelihood.
If we jump up and to the right from one gray curve to another, the population likelihood increases.
These constant-population-likelihood, hyperbolic curves do not depend on any population model. 
The population likelihood is then maximized by finding the point on the boundary of the hull tangent to the gray curve with the largest population likelihood (the most up-and-to-right gray curve).
In general, the maximum population likelihood point lies on the boundary of the hull~\citep{Silvey1980, Lindsay1983}. 
Our maximization problem can therefore be rewritten as a geometry problem.

We now turn our attention to the different rows of Fig.~\ref{fig:hull}.
In the top row, the two single-event likelihoods (right) are widely separated.
The cyan dot on the left-hand plot shows the maximum population likelihood point on the surface of the hull.
This is where the population likelihood has a value of \Lstroke.
It falls on a straight black surface of the hull, but not on the red atomic likelihood vector curve.
This means that the cyan point is a linear combination of two atomic likelihood vectors, which are indicated by the two black points (cf. Fig. 1 in Ref.~\citep{Lindsay1983}).
Thus, the maximum population likelihood solution consists of two delta functions, each corresponding to a different atomic vector.
This linear combination of delta functions is shown in the right-hand panel with cyan spikes.
Unsurprisingly, they coincide with the two single-event likelihood function peaks.

Moving down to the second row, the single-event likelihood functions (right) are now closer together.
The shape of the hull changes accordingly (left).
The hull boundary point that maximizes the population likelihood still does not fall on the red curve of atomic vectors.
Again, it is a linear combination of two black points.
However, since the shape of the hull has changed, the black points have moved relative to the top row.
The corresponding delta function spikes (right) therefore shift toward $\theta=0$ and no longer correspond to the maximimum likelihood points of the single-event likelihoods.

In the bottom row, the single-event likelihood functions (right) are closer still.
The hull (left) has now changed shape so that the cyan point marking the maximum population likelihood falls on the red curve denoting the set of atomic vectors (left).
This means that the likelihood can be maximized with a single delta function at $\theta=0$ (right).
In each case (and almost all scenarios, see Sec.~\ref{sapp:pathological}) the convex hull is unique, and so the cyan point of maximum population likelihood is unique as well. 
In all but the most pathological cases, Carath\'eodory's theorem~\citep{Caratheodory, roberts1973} states that all points on the boundary of a convex hull can be constructed by, at most, $N$ points that were used to initially construct the hull (in our problem these are the atomic likelihood vectors).
The relative weight of each delta function corresponds to the position along the boundary of the hull~\citep{Lindsay1983}. 
Thus, the population prior corresponding to the maximum population likelihood is a construction of a finite set of, at most, $N$ delta functions. 

The transition from two delta functions to one delta function occurs when the red curve passes through the black one (when the set of atomic likelihood vectors becomes convex).
During this transition, the cyan point changes from residing on a straight line connecting two atomic vectors to residing on a single atomic vector point.
This picture generalizes to higher dimensions.
Solutions with three delta functions (which can only exist when $N\geq3$) reside on two-dimensional planes.
Solutions with four delta delta functions (which can only exist when $N\geq4$) reside on three-dimensional hyper-planes.
And so on.

\subsection{Pathological cases}~\label{sapp:pathological}

\begin{figure*}
    \centering
    \includegraphics[width=0.86\linewidth]{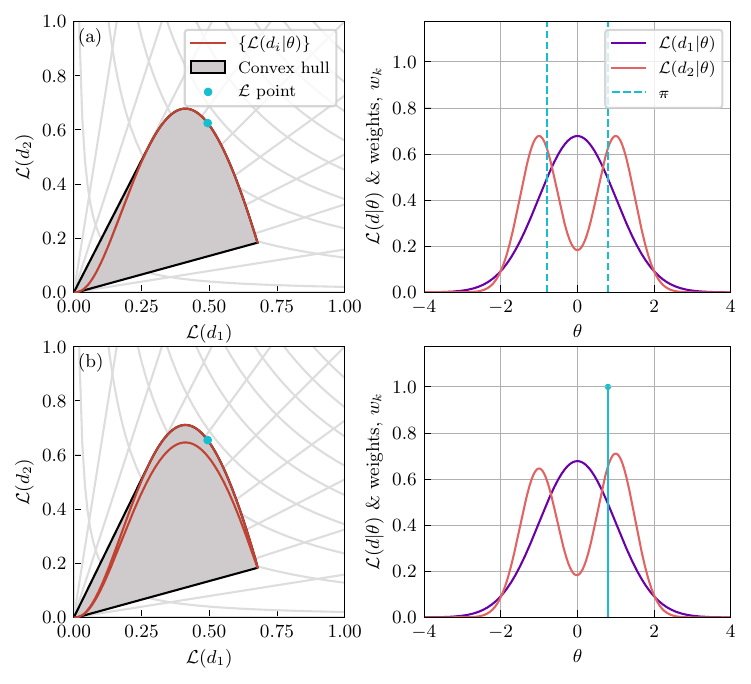}
    \caption{
    Demonstration of a pathological failure of the uniqueness of $\pistroke$. 
    This occurs when multiple distributions map to exactly the same point on the convex hull. 
    In (a), a perfectly symmetric, bimodal single-event likelihood has two delta functions with produce the same population likelihood. 
    Therefore, any combination of the two is a valid $\pistroke$. 
    However, such perfectly symmetric multi-modal distributions do not typically occur in gravitational-wave data analysis. 
    We see here we can break this degeneracy by only slightly breaking the symmetry, shown in (b).
    }
    \label{fig:hull_bad}
\end{figure*}

While we see that the maximum population likelihood almost always corresponds to a finite, unique set of $N$ or fewer delta functions, there are pathological cases (not likely to come up in real-world data analysis) where this is not the case. 
Such cases stem from the maximum population likelihood point not being unique.
So while the maximum population likelihood point is still found, multiple distributions can map to the same point in likelihood space. 
This requires artificial degeneracies in the measurements. 
In Fig.~\ref{fig:hull_bad}, we demonstrate one such example with two likelihood functions perfectly symmetric about $\theta=0$ and one of which is bimodal. 
In the likelihood space, the $\Lstroke$ point corresponds to two possible positions of the delta function. 
However, unlike in Fig.~\hyperref[fig:hull]{\ref{fig:hull}(a)} where the two possible delta function positions are separated, here they correspond to same point in likelihood space.
Therefore, any normalized combination of the two delta functions produces the maximum population likelihood.
This is emphasized by the dashed blue lines in the right column of Fig.~\hyperref[fig:hull_bad]{\ref{fig:hull_bad}(a)}, indicating that any combination of the two delta functions here is a permissible solution. 
However, we emphasize that this pathology arises from an artificial degeneracy, which is immediately broken if the likelihood functions are not precisely symmetric as demonstrated in Fig.~\hyperref[fig:hull_bad]{\ref{fig:hull_bad}(b)}.
Other, even more pathological, situations can be constructed where infinitely many atomic likelihood vectors reside at the maximum population likelihood point, allowing for arbitrarily structured $\pistroke$ distributions. 
However, all such situations require regions of perfectly uniform likelihood functions, which we do not expect in realistic observations---at least, not in gravitational-wave astronomy. 

\bibliography{emp}

\end{document}